\documentclass[twocolumn,preprintnumbers,amsmath,amssymb,superscriptaddress]{revtex4}
\usepackage{graphicx}
\usepackage{dcolumn}
\usepackage{bm}
\usepackage{soul}
\usepackage{color}
\usepackage{epstopdf}
\usepackage[version=3]{mhchem}
\usepackage{lipsum}
\usepackage[outercaption]{sidecap}
\usepackage{floatrow}
\usepackage{hyperref}

\begin{document}

\title{Structural Relaxation Time and Dynamic Shear Modulus of Glassy Graphene}
\author{Tran Dinh Cuong}
\email{cuong.trandinh@phenikaa-uni.edu.vn}
\affiliation{Phenikaa Institute for Advanced Study (PIAS), Phenikaa University, Hanoi 12116, Vietnam}
\author{Anh D. Phan}
\email{anh.phanduc@phenikaa-uni.edu.vn}
\affiliation{Phenikaa Institute for Advanced Study (PIAS), Phenikaa University, Hanoi 12116, Vietnam}
\affiliation{Faculty of Information Technology, Materials Science and Engineering, Artificial Intelligence Laboratory,  Phenikaa University, Hanoi 12116, Vietnam}
\affiliation{Department of Nanotechnology for Sustainable Energy, School of Science and Technology, Kwansei Gakuin University, Sanda, Hyogo 669-1337, Japan}
\author{Katsunori Wakabayashi}
\affiliation{Department of Nanotechnology for Sustainable Energy, School of Science and Technology, Kwansei Gakuin University, Sanda, Hyogo 669-1337, Japan}
\author{Pham Thanh Huy}
\affiliation{Phenikaa Institute for Advanced Study (PIAS), Phenikaa University, Hanoi 12116, Vietnam}
\affiliation{Phenikaa Research and Technology Institute,  Phenikaa University, Hanoi 12116, Vietnam}
\date{\today}

\date{\today}

\begin{abstract}
We theoretically investigate glass transition behaviors of the glassy graphene in a wide range of temperature, where this amorphous graphene is described as a hard-sphere fluid. The dynamic arrest of a particle is assumingly caused by interactions with its nearest neighbors and surrounding fluid particles. The assumption allows us to analyze roles of local and collective particle mobility. We calculate the temperature dependence of structural relaxation time and dynamic shear modulus, the dynamic fragility, and the glass transition temperature. In addition, correlations between these physical quantities are comprehensively discussed. Our theoretical calculations agree quantitatively well with recent simulations and Dyre's shoving model.

\end{abstract}

\maketitle


\section{INTRODUCTION}
Graphene is of importance in the development of modern science and technology due to its peculiar properties. Several representative applications are electronics \cite{1}, sensors \cite{2}, nanocomposites \cite{3}, energy storage \cite{4}, biomedical technologies \cite{5}, and membranes \cite{6}. Conventionally, graphene is viewed as a crystalline thin film having a honeycomb lattice structure. However, two recent simulations \cite{7,8} revealed the glass transition in monolayer and crumpled/disordered graphene. These studies showed that when liquid graphene is cooled down at a fast rate, carbon atoms are highly disordered and fall into amorphous states. The conformational predictions are qualitatively consistent with experimental observations of disordered graphene \cite{9,10}. At high temperatures ($\geq$ 1600 K), the three-dimensional (3D) glassy graphene, so-called graphene melt, behaves as a supercooled linear polymer melt but its  thermal stability is still much higher than ordinary polymers \cite{8}. The graphene melt can become a potential candidate material for working at extreme high-temperature regions. At low temperatures ($\leq 1600$ K), the 3D glassy graphene, known as "the graphene foam", undergoes a crossover from supercooled liquid to glassy state \cite{8}. The graphene foam can be used as a great filler in composite systems to overcome drawbacks caused by the dispersion and restacking of graphene \cite{11}. Moreover, physicochemical attributes of the graphene foam effectively reinforce multifunctional behaviors of composite materials \cite{12,13}.

The temperature dependence of structural relaxation time and dynamic shear modulus of bulk glassy graphene has been partially determined using coarse-grained molecular dynamics simulations in Ref.\cite{8}. The relaxation process of the graphene melt grows enormously with cooling up to 1 ns. Then, the simulation data was fitted and extrapolated to reach larger timescales (up to 100 s). Thus, it is very hard to capture computationally the structural dynamics of the 3D glassy graphene at low temperatures or experimental timescale \cite{8,14}. 

Recently, the Elastically Collective Nonlinear Langevin Equation (ECNLE) theory has been developed to quantitatively understand experimental data \cite{15}. The relaxation of a molecule is governed by its cage-scale dynamics, which is due to the nearest neighbor interactions, and long-range collective effects of fluid surroundings \cite{15}. Based on the basis of the treatment, we can predict the structural relaxation process over 14 decades in time \cite{15}. The ECNLE theory has successfully described colloidal suspensions \cite{15,16}, supercooled molecular liquids \cite{16,17}, amorphous drugs \cite{18,36}, and polymer melts \cite{19}. However, it has not yet been applied to study dynamics properties of glassy graphene. 

In this paper, we apply the ECNLE theory to determine the glass transition behaviors of the 3D glassy graphene \cite{8}. Effects of temperature on the structural relaxation time and the dynamic shear modulus are theoretically calculated. Our calculations are compared to the simulation data of Ref.\cite{8}. Moreover, we focus on correlations between the shear response and the activated hopping process. The obtained results show a good agreement with Dyre's shoving model \cite{20,21}.

\section{THEORETICAL BACKGROUND}
\begin{figure}[h]
\includegraphics[width=9 cm]{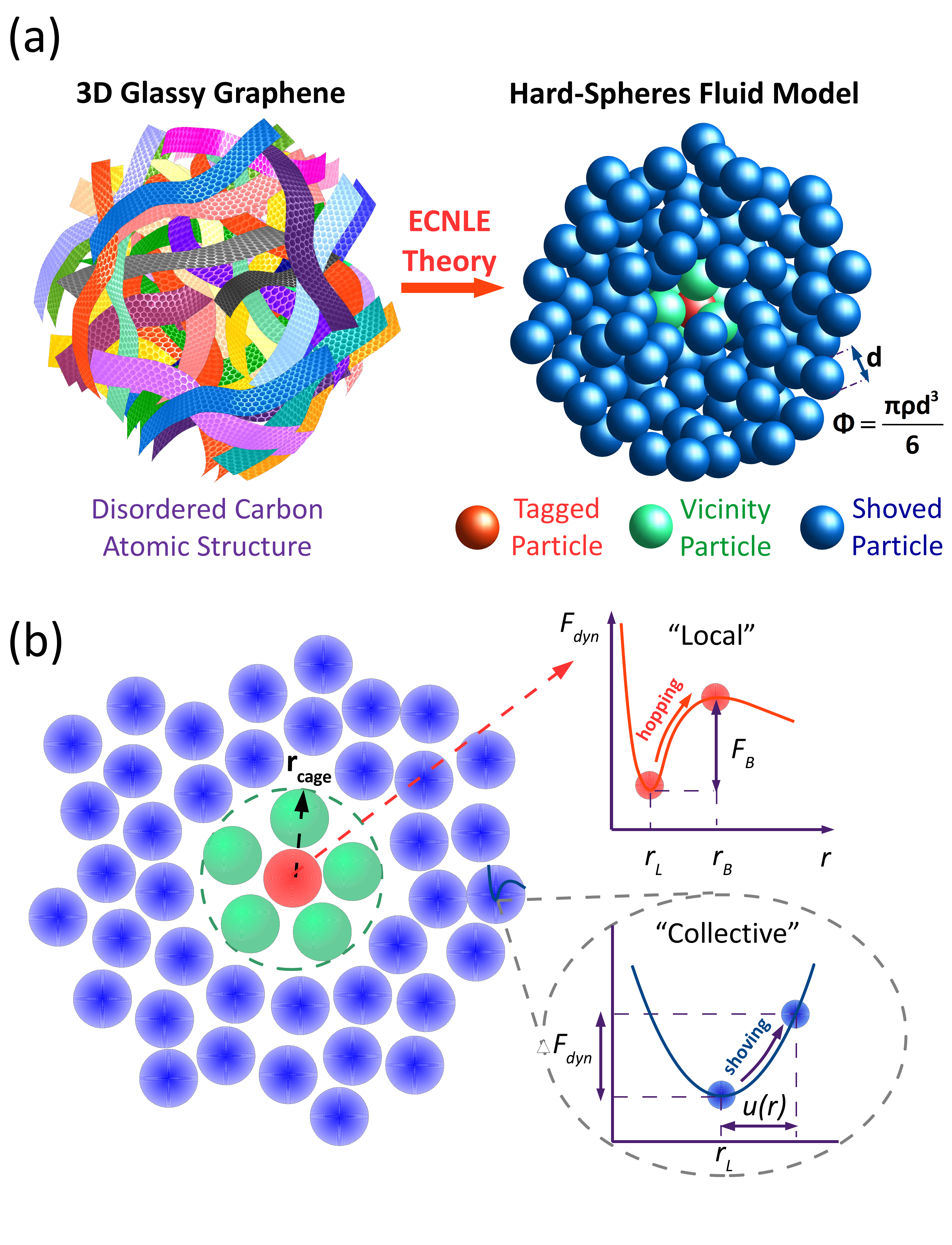}
\caption{\label{fig:1}(Color online) (a) Illustration of a disordered structure for the 3D glassy graphene \cite{8} and its modeling based on hard-spheres fluids. (b) Schematic of the activated hopping process described via the dynamic free energy leading to long-range harmonic vibrations in surroundings.}
\end{figure}

Simulations in Ref.\cite{8} suggest that packing a large number of crumpled and disoriented graphene nanosheets into a random configuration (as shown in Figure \ref{fig:1}a) can form a linear polymer-like structure. This disordered structure can be described as an assembly of impenetrable spherical particles having the particle diameter, $d$, the average number density of particles, $\rho$, and the volume fraction, $\Phi=\pi\rho d^3/6$ in the framework of the ECNLE theory \cite{15,16,17,18,19,36}. Here, we assume that each carbon atom corresponds to an effective hard-sphere particle in our ECNLE model. Although the binding energy between two nearest atoms is strong in the crystal form, this interaction is significantly weakened in the glassy state. To zeroth-order approximation, effects of interparticle interactions on the molecular mobility are only encoded in the density-to-temperature conversion or the thermal expansion process, which will be presented later in this work. Recall that the treatment has been successfully applied to investigate various amorphous drugs, polymers, and thermal liquids  \cite{15,16,17,18,19,36}.

Based on the Percus-Yevick (PY) theory \cite{25}, we obtain structural information for a hard-sphere fluid including the radial distribution function, $g(r)$, the static structure factor, $S(q)$, and the direct correlation function, $C(q)$, where $q$ is the wave vector. It is well-known that PY calculations agree well with simulations \cite{25}. Based on PY theory, the direct correlation function in the real-space is
\begin{eqnarray}
C(r)&=&-\frac{(1+2\Phi)^2}{(1-\Phi)^4}+\frac{6\Phi(1+\Phi/2)^2}{(1-\Phi)^4}\frac{r}{d}\nonumber\\
&-&\frac{\Phi(1+2\Phi)^2}{2(1-\Phi)^4}\left(\frac{r}{d}\right)^3\quad for\quad r\leq d,\\
C(r)&=&0\quad for\quad r>d.
\label{eq:1}
\end{eqnarray}
Using the Fourier transform gives
\begin{eqnarray}
C(q)&=&\frac{4\pi}{q}\int_0^dC(r)\sin(qr)rdr,\\
S(q)&=&\frac{1}{1-\rho C(q)},\\
g(r)&=&1+\frac{1}{2\pi^2\rho r}\int_0^\infty\left[S(q)-1\right]q\sin(qr)dq.
\label{eq:1}
\end{eqnarray}

The mobility of a single particle is captured in a picture of slow dynamics. In the overdamped limit, the nonlinear Langevin equation describing the particle motion is \cite{26,27}
\begin{eqnarray}
\zeta_s\frac{\partial r}{\partial t}-\frac{\partial F_{dyn}}{\partial t}+\delta f=0,
\label{eq:1}
\end{eqnarray}
where $r$ is the scalar displacement of the tagged particle, $\zeta_s$ is the short-time friction constant, $\delta f$ is the noise random force obeying the Gaussian correlation function $\left<\delta f(0)\delta f(t)\right>=2k_BT\zeta_s\delta(t)$, and $F_{dyn}(r)$ is the dynamic free energy quantifying effects of the nearest neighbors on the particle. An analytical expression of $F_{dyn}(r)$ is \cite{26,27}
\begin{eqnarray}
\frac{F_{dyn}}{k_BT}&=&-3\ln\frac{r}{d}-\int{\frac{d\textbf{q}}{(2\pi)^3}\rho C^2(q)S(q)}\left[1+S^{-1}(q)\right]^{-1}\nonumber\\
&\times& \exp{\left\{-\frac{q^2r^2}{6}\left[1+S^{-1}(q)\right]\right\}},
\label{eq:2}
\end{eqnarray}
where $k_B$ is Boltzmann's constant and $T$ is temperature. The above equation contains two terms corresponding to the ideal fluid state (the first term) and the dynamic mean-field trapping potential (the second term) \cite{26,27}. In dilute solutions, fluid particles are not localized since the trapping force is weak. When $\Phi\geq 0.432$, the confinement effects become stronger than delocalizing tendency. Thus, the tagged particle is dynamically arrested within a local cage formed by its nearest neighbors \cite{26,27}. The cage radius, $r_{cage}$, is estimated as the first minimum of $g(r)$ \cite{15,16,17,18,19,36}. In the framework of the PY theory \cite{25}, $r_{cage}$ merely changes from $1.52d$ to $1.28d$ when $0.44\leq\Phi\leq0.64$. For simplicity, one can approximate $r_{cage}\approx1.5d$ \cite{15,16,17,18,19,36}.

As shown in Figure \ref{fig:1}b, the dynamic energy profile provides key quantities in the cage-scale dynamics. These physical quantities are the localization length, $r_L$, the barrier position, $r_B$, and the local barrier height, $F_B=F_{dyn}(r_B)-F_{dyn}(r_L)$. By associating the analysis with the Green-Kubo formula \cite{28,29}, we can calculate the dynamic shear modulus by 
\begin{eqnarray}
G=\frac{k_BT}{60\pi^2}\int_{0}^{\infty}dq\left[\frac{q^2}{S(q)}\frac{\partial S(q)}{\partial q}\right]^2e^{-q^2r^2_L/3S(q)}.
\label{eq:3}
\end{eqnarray}
At high densities, the localization length is significantly reduced, and particles are highly-localized. In this limit, only high wavelengths contribute to the dynamic shear modulus \cite{30}. After straightforward analysis, Equation \eqref{eq:3} can be re-written as
\begin{eqnarray}
G=\frac{9\Phi}{5\pi}\frac{k_BT}{dr_L^2}.
\label{eq:4}
\end{eqnarray}

To allow a large hop out of the cage, the activated hopping causes a cage expansion and cooperative reorganization of surrounding particles. The fluctuation on the surface of the particle cage excites other particles in medium to vibrate. The oscillation is assumed to be harmonic and generates a displacement field, $u(r)$, outside the cage. Thus, one can analytically formulate $u(r)$ by continuum mechanics analysis \cite{31}
\begin{eqnarray}
u(r)=\Delta{r_{eff}}\left(\frac{r_{cage}}{r}\right)^2, \quad r \geq r_{cage}
\label{eq:5}
\end{eqnarray}
where $\Delta{r_{eff}}$ is an average cage expansion amplitude calculated by $\Delta{r_{eff}}\approx\frac{3\left(r_B-r_L\right)^2}{32r_{cage}} \ll r_L$ \cite{15,16,17,18,19,36}. 

According to Einstein's glass model, a harmonic oscillation of a fluid particle is equivalently considered as a spring pendulum having spring constant $K_L=\left[\partial^2{F_{dyn}(r)}/\partial{r^2}\right]_{r=r_L}$. Consequently, the elastic energy stored in the surrounding medium is written by \cite{15,16,17,18,19,36}
\begin{eqnarray}
F_{E}&=&\int_{r_{cage}}^{\infty}{\frac{1}{2}K_Lu^2(r)\rho g(r)4\pi r^2dr}\nonumber\\
 &=&12\Phi\Delta{r_{eff}^2}\left(\frac{r_{cage}}{d}\right)^3K_L,
\label{eq:6}
\end{eqnarray}
where we approximate $g(r)\approx 1$ when $r \geq r_{cage}$. In the deeply supercooled regime (high densities), one can analytically relate $F_E$ to $G$ by \cite{15,16,17}
\begin{eqnarray}
F_{E}=20\pi\frac{\Delta r_{eff}^2r_{cage}^3}{d^2}G.
\label{eq:7}
\end{eqnarray}

The activated hopping process is now affected by both cage-scale and collective dynamics which are characterized by the local and elastic barriers. These two barriers depend on the volume fraction. A role of the elastic barrier becomes important in the relaxation event at high densities $\left(\Phi\geq 0.54\right)$ \cite{15,16,17,18,19,36}. When $\Phi\geq 0.6$, a growth of the elastic barrier with increasing the density is more than that of the local analog. The presence of the elastic barrier gives rise to non-Arrhenius behaviors in the glass-forming liquids \cite{15,16,17,18,19,36}. The structural relaxation time, $\tau_\alpha$, is known as the mean time for a tagged particle to escape from its cage. This can be computed using the modified Kramer theory as \cite{15,16,17,18,19,36}
\begin{eqnarray}
 \tau_\alpha=\tau_s\left[1+\frac{2\pi}{\sqrt{K_LK_B}}\frac{k_BT}{d^2}e^{F_{total}/k_BT}\right],
\label{eq:8}
\end{eqnarray}
where $F_{total}=F_E+F_B$ is the total barrier, $\tau_s$ is the short relaxation timescale and $K_B=-\left[\partial^2{F_{dyn}(r)}/\partial{r^2}\right]_{r=r_B}$ is the absolute curvature at the barrier position. The explicit form of $\tau_s$ can be found in a previous work \cite{18}. Now, the structural relaxation time can be obtained as a function of the volume fraction $\Phi$. 

To compare between the ECNLE calculation and experiment, one requires a density-to-temperature conversion (thermal mapping) to include the thermal effects on the structural relaxation. Phan and his co-workers \cite{18} proposed a simple thermal mapping given by
\begin{eqnarray}
\Phi\approx\Phi_0\left[1-\beta\left(T-T_0\right)\right],
\label{eq:9}
\end{eqnarray}
where $\Phi_0=0.5$, $\beta$ is the volume thermal expansion coefficient of the glassy graphene, and $T_0=T_{ref}+\left(\Phi_{ref}-\Phi_0\right)/\beta\Phi_0$ is a characteristic temperature which depends on material-specific details. For simplicity, the reference temperature $T_{ref}$ is taken from experiments or simulations, then $\Phi_{ref}$ is identified by synchronizing $\tau_\alpha(\Phi_{ref})$ and $\tau_\alpha(T_{ref})$ obtained from the ECNLE theory and references, respectively \cite{18}. By using ECNLE calculations, we find $\tau_\alpha(\Phi=0.5604)=0.86$ ns which is equivalent to $\tau_\alpha(T=2495.68$ K$)= 0.86$ ns in the simulation in Ref.\cite{8}. The crossing enables to estimate $\Phi_{ref}=0.5604$ and $T_{ref}=2495.68$ K.

In Ref.[8], authors revealed some analogies between the glass transition processes in the 3D glassy graphene and linear chain polymers. Hence, one can roughly estimate the volume thermal expansion coefficient of the 3D glassy graphene in the supercooled state via a simple relation, which is \cite{32} 
\begin{eqnarray}
\beta T_g\approx 0.16\pm 0.03,
\label{eq:10}
\end{eqnarray}
where $T_g$ is the glass transition temperature obeying $\tau_\alpha(T_g)=100$ s. Applying $T_g=1600$ K \cite{8} gives $\beta= 8-12 \times 10^{-5}$ K$^{-1}$. On the other hand, the 3D glassy graphene is also a variant form of two-dimensional substance, so its volume thermal expansion coefficient is expected to range from 4 to $8 \times 10^{-5}$ K$^{-1}$ at 1600 K \cite{33,34}. Taking the intersection of two mentioned sets, we have $\beta=8\times 10^{-5}$ K$^{-1}$. This reasonable value is supported by recently reported data for several glass-forming liquids having $T_g>1000$ K \cite{35}. 

In principle, when a supercooled liquid falls out of equilibrium, its volume thermal expansion coefficient may be abruptly reduced \cite{32}. Thus, the growth rate of the structural relaxation time with cooling near $T_g$ may decrease anomalously \cite{36}. However, this distinctive event, so-called a dynamic structural decoupling, does not occur in all glass-forming liquids. For example, in the case of chloramphenicol \cite{37} and polystyrene \cite{38}, the decoupling does not occurs when $\tau_\alpha$ increases up to $10^3$ s. In ECNLE calculations, it implies that the thermal expansion coefficients of chloramphenicol and polystyrene remain unchanged above and below $T_g$ \cite{36}. Although the decoupling phenomenon has been under debate for decades, its underlying mechanisms is still ambiguous \cite{39}. Especially, for unexplored systems such as glassy graphene, there are no experiment, simulation, and theory to suggest how the decoupling tendency takes place in the glassy state. Thus, in this study, we approximately assume that the graphene melt and the graphene foam have the same thermal expansion coefficient of $\beta=8\times 10^{-5}$ K$^{-1}$.

\section{RESULTS AND DISCUSSION}

\begin{figure}[htp]
\includegraphics[width=9 cm]{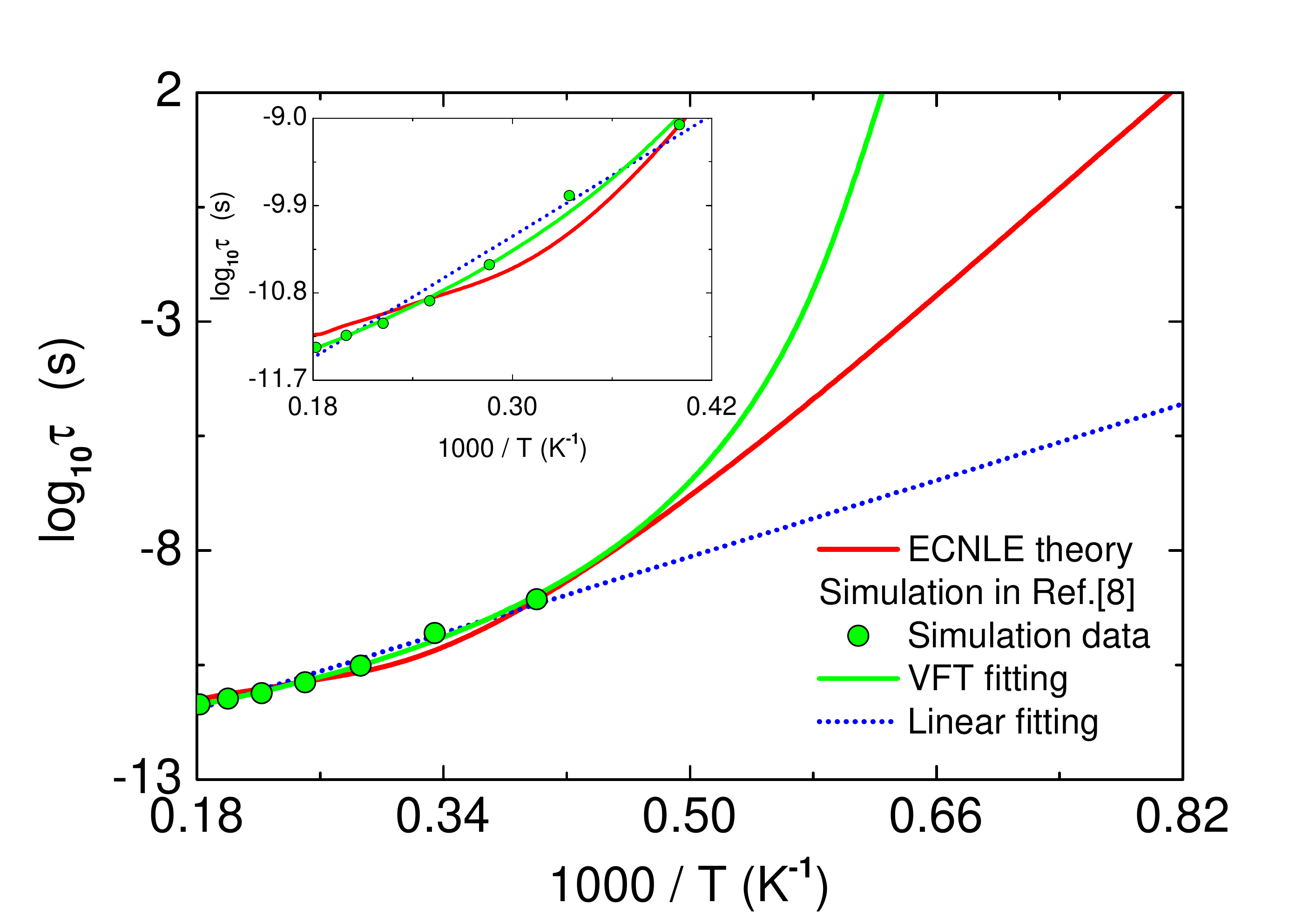}
\caption{\label{fig:2}(Color online) Effects of temperature on the logarithm of the structural relaxation time calculated using the ECNLE theory and molecular dynamics simulation in Ref.\cite{8}. Simulation data is fitted by Equation (\ref{eq:11}) \cite{40,41,42} and linear regression analysis. Inset: The same plot as the mainframe but the simulation timescale is zoomed in.}
\end{figure}

Figure \ref{fig:2} shows the logarithm of the structural relaxation time of the 3D glassy graphene as a function of inverse normalized temperature. One can realize that $\log_{10}\tau_\alpha$ increases dramatically with cooling and this variation exhibits a non-Arrhenius behavior. The temperature sensitivity of $\log_{10}\tau_\alpha$ can be quantitatively described by the Vogel-Fulcher-Tammann (VFT) equation as \cite{40,41,42}
\begin{eqnarray}
\log_{10}\tau_\alpha=\log_{10}\tau_0+\frac{DT_{VFT}}{T-T_{VFT}},
\label{eq:11}
\end{eqnarray}
where VFT fitting parameters $\log_{10}\tau_0$, $D$, and $T_{VFT}$ are presented in Table \ref{table:1}. It is clear to see that our numerical calculations are consistent well with simulation \cite{8} at high temperatures. Nevertheless, at the deeply supercooled regime, the difference between these methods becomes apparent. For example, combining Equation (\ref{eq:8}) and (\ref{eq:9}) gives $T_g=1230$ K. While extrapolating the VFT fit function for simulation data \cite{8} up to 100 s provides $T_g=1600$ K. The obtained values for $T_g$ exhibit uncertainty because simulation timescale cannot exceed the nanosecond domain and capture collective physics \cite{8}. Another reason for the presence of deviation is that the VFT fitting is a gross extrapolation. One can find the deviation even when the time-scale data is in a deeply supercooled regime \cite{36}. Furthermore, the high rigidity and porous nature related to the packing of distorted graphene nanosheets may have a significant impact on the glass transition \cite{8}. However, overall, the glass transition temperature of the 3D glassy graphene is still exceptionally higher compared to that of ordinary amorphous materials \cite{18,32}. This conclusion indicates that the 3D glassy graphene has great potential for high-temperature applications due to its excellent thermal stability \cite{8}.

\begin{table}[h!]
\centering
\begin{tabular}{|c| c| c| c| c|} 
 \hline
Method & $\log_{10}\tau_0$ & $DT_{VFT}$ (K) & $T_{VFT}$ (K) \\ 
 \hline
ECNLE theory & -13.3 & 7920.5 & 750.1 \\ 
Simulation \cite{8} &-12.3 & 3883.4 & 1328.5 \\
 \hline
\end{tabular}
\caption{The VFT fitting parameters for the 3D glassy graphene obtained from the ECNLE theory and simulation in Ref.\cite{8}.}
\label{table:1}
\end{table}

The glass transition behaviors of 3D glassy graphene are crucial for the mechanical properties \cite{20,21,43}. At low temperatures (near and far below $T_g$), in the framework of the ECNLE theory, the collective elastic barrier dominates the local barrier in the relaxation process. Moreover, Equation (\ref{eq:7}) suggests that $F_E$ grows dramatically with increasing the dynamic shear modulus \cite{15,16,17}. This is qualitatively consistent with a shoving model proposed by Dyre and his co-workers \cite{20,21}. 

To clarify the correlation between the thermally activated hopping event and elasticity data, we compare our results with the (Dyre's) shoving model. An analytical expression of the temperature dependence of the relaxation time in this shoving model is \cite{20,21}
\begin{eqnarray}
\tau_\alpha(T) = \tau_c\exp\left(\frac{GV_c}{k_BT}\right),
\label{eq:12}
\end{eqnarray}
where $V_c$ is a characteristic volume and $\tau_c$ is the relaxation time at a high temperature limit (far above $T_g$). In many works \cite{20,21}, it is assumed that $V_c$ is independent of temperature. From this assumption, one obtains
\begin{eqnarray}
\log_{10}\tau_\alpha(T) &=& \left[\log_{10}\frac{\tau_\alpha(T_g)}{\tau_c}\right]\frac{G(T)T_g}{G(T_g)T}+\log_{10}\tau_c \nonumber\\
&=& \left[\log_{10}\frac{\tau_\alpha(T_g)}{\tau_c}\right]\chi(T)+\log_{10}\tau_c.
\label{eq:13}
\end{eqnarray}
where $\chi(T)=T_gG(T)/TG(T_g)$ is a dimensionless response function of shear modulus. Since $\tau_\alpha(T_g)=100$ s by definition, Equation (\ref{eq:13}) reveals a linear relationship between $\log_{10}\tau_\alpha(T)$ and $\chi(T)$. 

\begin{figure}[htp]
\includegraphics[width=9 cm]{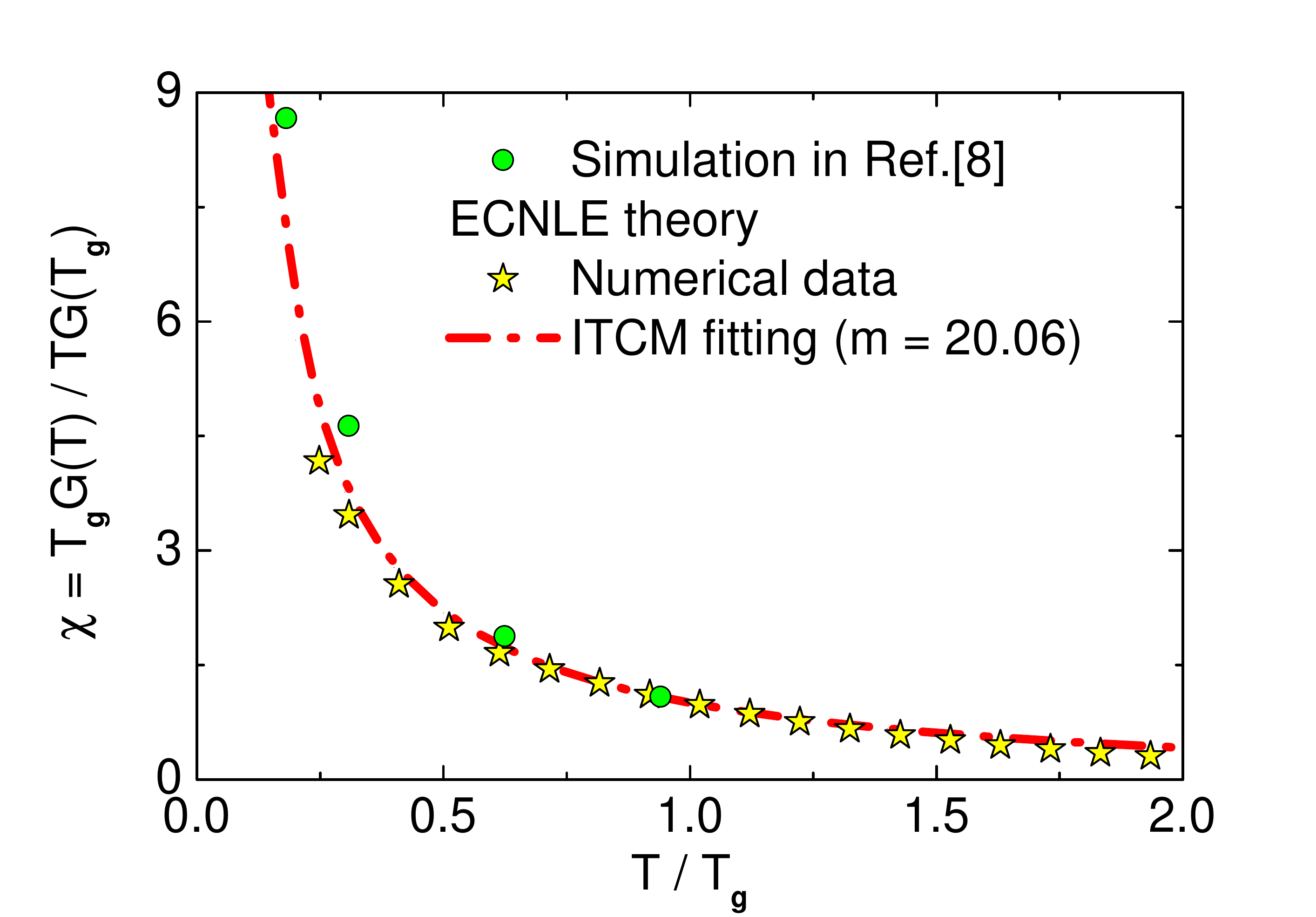}
\caption{\label{fig:3}(Color online) Variation of $\chi=T_gG(T)/TG(T_g)$ as a function of $T/T_g$ computed using the ECNLE theory, ITCM fitting \cite{44,45,46} and molecular dynamics simulation data of Ref.\cite{8}.}
\end{figure}

Figure \ref{fig:3} shows the temperature dependence of susceptibility $\chi(T)$ calculated by ECNLE theory together with the simulation data of Ref.\cite{8}. It is clear to see that the $\chi(T)$ decreases with increasing $T/T_g$. This is because the heating process accelerates molecular mobility and weakens binding among fluid particles. By combining the ultra-local analysis (Equation (\ref{eq:4})) \cite{30} and an exponential law $r_L=30d\exp(-12.2\Phi)$ \cite{26}, one obtains
\begin{eqnarray}
\chi=\frac{\Phi e^{24.4\Phi}}{\Phi_ge^{24.4\Phi_g}},
\label{eq:14}
\end{eqnarray}
where $\Phi_g \approx 0.6110$. Equation (\ref{eq:14}) clearly explains why $\chi(T)$ monotonically decreases with $T/T_g$ even around $T = T_g$.

Since our theoretical approach cannot access the temperature below $0.24T_g$ due to limitations of the thermal mapping, we need to find another approach to investigate molecular dynamics at lower temperatures. In a recent work \cite{20}, Dyre and his co-worker compared various functional forms of the dynamic shear modulus to experimental results for glycerol and DC704. Notably, they pointed out that the Interstitialcy Theory of Condensed Matter (ITCM) \cite{44,45,46} can accurately describe the elasticity data over a wide range of temperatures as
\begin{eqnarray}
\chi=\frac{T_g}{T}\exp\left[\gamma\left(1-\frac{T}{T_g}\right)\right],
\label{eq:15}
\end{eqnarray}
where $\gamma = m/17 - 1$ with the dynamic fragility $m$ defined by
\begin{eqnarray}
m=\left[\frac{d\log_{10}\tau_\alpha}{d(T_g/T)}\right]_{T=T_g}.
\label{eq:16}
\end{eqnarray}
In a physical picture of the ITCM, the activation energy is close to the interstitialcy diffusion energy, which is proportional to the dynamic shear modulus \cite{44,45,46}. These findings are consistent with continuum mechanics analysis of Dyre and his co-workers \cite{20,21}. Interestingly, one can decude the VFT relation (Equation (\ref{eq:11})) by combining Equations (\ref{eq:12}) and (\ref{eq:15}) \cite{46}. This analysis also reveals a microscopic viewpoint underlying the VFT equation. Hence, it is reasonable to capture the mechanical behaviors of the 3D glassy graphene via the ITCM.

As shown in Figure \ref{fig:3}, applying the ITCM fit function to our numerical data gives $\gamma=0.18$ or $m=20.06$. This value quantitatively agrees with $m=23.4$ predicted by the ECNLE theory. Thus, the 3D glassy graphene is a strong material $(m<30)$. Furthermore, our ECNLE calculations, ITCM fitting function, and simulation data of Ref.\cite{8} are close to each other. This agreement validates the application of ECNLE theory to the 3D glassy graphene.

\begin{figure}[htp]
\includegraphics[width=9 cm]{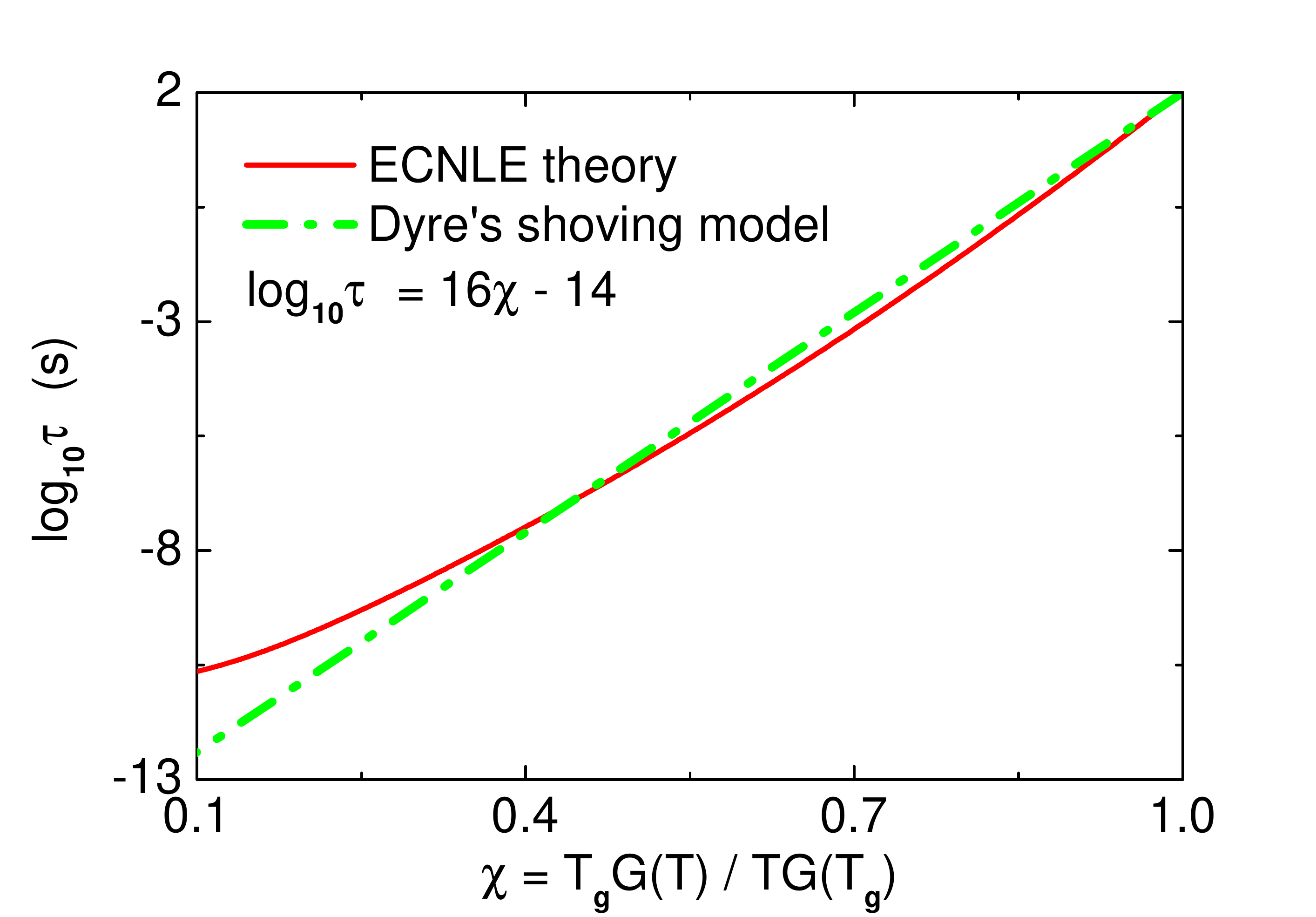}
\caption{\label{fig:4}(Color online) The logarithm of the structural relaxation time versus $\chi$ calculated using the ECNLE theory and Dyre's shoving model \cite{20,21}.}
\end{figure}

By fitting Equation (\ref{eq:13}) with experimental data of polymers and thermal liquids \cite{20,21}, Dyre found a simple function: $\log_{10}\tau_\alpha=16\chi-14$. Figure \ref{fig:4} shows logarithmic plot of $\tau_\alpha$ on the parameter $\chi$ for glassy graphene obtained from the ECNLE theory, together with the Dyre's fitting function. Quantitative predictions given by these two approaches are close in the deeply supercooled regime. When $0.4\leq\chi\leq1$, since $F_E\propto F_B^2$, $F_B\propto 1/r_L$ and $1/r_L^2\propto G/(T\Phi)$ \cite{16,17}, we can analytically analyze $\log_{10}\tau_\alpha \propto F_{total}/k_BT$ as a funtion of $\chi$
\begin{eqnarray}
\frac{F_{total}}{k_BT}=a_1\frac{\chi}{\Phi}+a_2\sqrt{\frac{\chi}{\Phi}}+a_3,
\label{eq:17}
\end{eqnarray}
where $a_1=48.09$, $a_2=-53.43$, and $a_3=22.12$ are fit parameters. This relation seems to be more complicated than the Dyre's harmonic form $F_{total}/k_BT\propto \chi$ \cite{20,21} because it captures physics of local interatomic interactions. The absence of cage-scale dynamics causes the failure of Dyre's shoving model in the simulation time scale ($\tau_\alpha<10^{-9}$ s) \cite{15,16,17}, while the ECNLE theory still works well in this region (as shown in Figure \ref{fig:2}). However, near the glass transition temperature, employing the Taylor expansion for $F_{total}/k_BT$ around $\chi_g=1$ gives us
\begin{eqnarray}
\frac{F_{total}}{k_BT}&=&\frac{a_2\Phi_g\sqrt{\Phi_g}+\left(2a_1+a_2\sqrt{\Phi_g}\right)a_4\chi_g}{2\Phi_g^2}\chi_g+\nonumber\\
&+&a_3+\frac{\left(2a_1+a_2\sqrt{\Phi_g}\right)\left(\Phi_g-a_4\chi_g\right)}{2\Phi_g^2}\chi,
\label{eq:18}
\end{eqnarray}
where $a_4=\left(d\Phi/d\chi\right)_{T=T_g}=0.038$. Equation (\ref{eq:18}) suggests that the linear-growth rule for $\log_{10}\tau_\alpha(\chi)$ has high accuracy when $0.8\leq \chi\leq 1$. Overall, understanding of the correlation between $\chi$ and $\tau_\alpha$ can suggest a way of determining molecular mobility via mechanical measurements \cite{20,21}.

\section{CONCLUSION}
The temperature dependence of structural relaxation and mechanical properties of the 3D glassy graphene has been investigated using the ECNLE theory. At the high temperature regime, the theoretical $\tau_\alpha(T)$ are quantitatively consistent with the previous simulation data of Ref.\cite{8}. At low temperatures, the relation between $\tau_\alpha(T)$ and the dynamic shear modulus $G(T)$ has been considered. We have calculated the response of $\chi=T_gG(T)/TG(T_g)$ to $T/T_g$ and $\log_{10}\tau_\alpha(T)$. Our numerical results are in good accordance with the ITCM fit equation, Dyre's shoving model, and simulation data of Ref.\cite{8}. In addition, the ITCM fitting gives the dynamic fragility of glassy graphene equal to 20.06, while the value predicted by our ECNLE calculations is 23.4. These agreements clearly validate our analytical approach for understanding glassy states and the dynamic shear modulus of unexplored systems. Furthermore, it is possible to apply the ECNLE theory to study the glassy dynamics in two-dimensional systems.

\begin{acknowledgements}
This work was supported by JSPS KAKENHI Grant Numbers JP19F18322 and JP18H01154.
\end{acknowledgements}

\end{document}